# ZeroFold: Protein-RNA Binding Affinity Predictions from Pre-Structural Embeddings

Josef Hanke[1], Sebastian Pujalte Ojeda[1], Shengyu Zhang[1], Werngard Czechtizky[2], Leonardo De Maria[2] and Michele Vendruscolo[1,*]

[1]*Centre for Misfolding Diseases, Yusuf Hamied Department of Chemistry, University of Cambridge, Cambridge, UK*
[2]*Medicinal Chemistry, Research and Early Development, Respiratory and Immunology, BioPharmaceuticals R&D, AstraZeneca, Gothenburg, Sweden*

## Abstract

The accurate prediction of protein-RNA binding affinity remains an unsolved problem in structural biology, limiting opportunities in understanding gene regulation and designing RNA-targeting therapeutics. A central obstacle is the structural flexibility of RNA, as, unlike proteins, RNA molecules exist as dynamic conformational ensembles. Thus, committing to a single predicted structure discards information relevant to binding. Here, we show that this obstacle can be addressed by extracting pre-structural embeddings, which are intermediate representations from a biomolecular foundation model captured before the structure decoding step. Pre-structural embeddings implicitly encode conformational ensemble information without requiring predicted structures. We build ZeroFold, a transformer-based model that combines pre-structural embeddings from Boltz-2 for both protein and RNA molecules through a cross-modal attention mechanism to predict binding affinity directly from sequence. To support training and evaluation, we construct PRADB, a curated dataset of 2,621 unique protein-RNA pairs with experimentally measured affinities drawn from four complementary databases. On a held-out test set constructed with 40% sequence identity thresholds, ZeroFold achieves a Spearman correlation of 0.65, a value approaching the ceiling imposed by experimental measurement noise. Under progressively fairer evaluation conditions that control for training-set overlap, ZeroFold compares favourably with respect to leading structure-based and leading sequence-based predictors, with the performance gap widening as sequence similarity to competitor training data is reduced. These results illustrate how pre-structural embeddings offer a representation strategy for flexible biomolecules, opening a route to affinity prediction for protein-RNA pairs for which no structural data exist.

\* Correspondence to: mv245@cam.ac.uk

**Introduction**

Protein-RNA interactions are fundamental to cellular regulation, and their misregulation is increasingly recognized as a driver of human disease[1-3]. Long-read sequencing technologies have revealed an unexpectedly diverse repertoire of non-coding RNAs with key regulatory functions across cell types and tissues[4,5], and the dysregulation of these molecules has been implicated in a wide range of diseases, from defective RNA splicing to aberrant microRNA activity[6,7]. Protein-RNA interactions are central to these processes and are equally relevant to the development of peptide-based therapeutics, a clinically validated drug class whose larger interaction surfaces make them well suited to the shallow, extended interfaces typical of RNA-binding sites[8,9]. Recent advances in understanding the protein-RNA interaction code, together with computational tools for predicting biomolecular structures and interactions[10-12], have created new opportunities for the rational design of RNA-binding peptides.

The greatest difficulty in predicting these interactions arises from the nature of the binding interfaces themselves. Protein-RNA complexes typically involve large, often shallow surface areas that lack the hydrophobic pockets characteristic of traditionally druggable targets[13], and the binding is frequently heterogeneous, involving disordered protein regions. Furthermore, the flexibility of RNA structures often causes them to behave like intrinsically disordered proteins, so that a single three-dimensional conformation cannot adequately represent the bound state[14].

Several computational strategies have been developed to address this problem, broadly categorised into structure-based and sequence-based approaches. Structure-based methods extract features at the binding interface, such as contact distances or interaction energies, from experimentally resolved structures and use them to train affinity prediction models. While these approaches can capture detailed physical interactions, they rely on structural inputs that are available for only a small fraction of known protein-RNA pairs and tend to show limited generalisation to novel sequences. Among these, PRdeltaGPred[15] and PredPRBA[16] both utilise interface structure features in addition to sequence information, and the recent model CoPRA combines protein and RNA language model representations with complex structural information, reporting strong performance on complexes with resolved structures[17]. Sequence-based models avoid this structural dependency but have shown variable performance. PNAB classifies protein-nucleic acid complexes by nucleic acid type and builds stacking heterogeneous ensemble models for each class[18], PRA-pred uses multiple linear regression[19], and the leading sequence-based model DeePNAP encodes sequences directly[20], although its performance has been shown to vary substantially depending on the test set[17]. A central limitation across all these approaches is the scarcity of high-quality training data, with existing datasets comprising only a few thousand data points.

Machine learning has transformed the prediction of protein structure, protein-protein interactions, and a growing range of biomolecular properties[10,21-23]. Yet accurately modelling biomolecules that exist as conformational ensembles, such as RNA and disordered proteins, remains a fundamental challenge[24,25]. Current structure prediction methods output a single conformation, discarding the ensemble information that is often critical for understanding binding. This limitation is especially acute for protein-RNA interactions, where the flexibility of both partners means that no single predicted structure can fully capture the determinants of affinity.

A key insight motivating this work is that the intermediate representations generated by biomolecular foundation models during structure prediction, here termed pre-structural embeddings, encode richer information about binding than the final predicted structures themselves. Rather than committing to a single folded conformation, these high-dimensional representations implicitly encode the ensemble of possible conformations and their associated binding properties[26,27], a distinction that is particularly consequential for flexible molecules such as RNA. Crucially, extracting these embeddings before the structure decoding step avoids the information loss that occurs when a dynamic conformational ensemble is collapsed into a single predicted structure. This principle has recently been validated for protein-ligand binding affinity prediction, where pre-structural embeddings achieved strong predictive performance[27], suggesting that it may represent a general strategy for modelling biomolecular interactions involving flexible partners.

Here we introduce ZeroFold, a transformer-based model that operationalises this principle for protein-RNA binding affinity prediction. ZeroFold extracts pre-structural embeddings from Boltz-2[28], a biomolecular foundation model that natively supports protein, RNA, and DNA sequences within a unified trunk architecture. Representations are extracted from the final trunk layer prior to the structure prediction head, capturing contextually rich per-residue and pairwise embeddings that reflect both sequence and evolutionary context (**Figure 1**). These are processed through separate encoding blocks for each chain, integrated via a cross-modal attention module, and passed to an affinity prediction head. The choice of Boltz-2 over protein-only foundation models was motivated by its unified treatment of nucleic acid and protein sequences, which makes it well suited to modelling the joint conformational space of protein-RNA complexes, and by its flexible sequence definitions, which accommodate non-natural amino acids and modified nucleotide bases relevant to therapeutic design.

To address the scarcity of training data, we assembled the Protein-RNA Affinity Database (PRADB), a curated dataset of 4,510 pairs (2,621 unique) of protein and RNA sequences with experimentally measured binding affinities ($pK_D$), compiled from four complementary sources: ProNAB[29], BioLiP2[30], the UTexas Aptamer Database[31], and PDBbind+[32]. We train ZeroFold on PRADB and evaluate it against state-of-the-art structure-based and sequence-based methods. On a test set constructed with strict 40% sequence identity thresholds on both protein and RNA sequences, ZeroFold achieves a Spearman correlation of 0.65, performing better than CoPRA and DeePNAP even in baseline comparisons. Because the ZeroFold test set contains sequences present in the training datasets of these competing models, this comparison inherently favours those models. To account for this, we constructed progressively stricter evaluation subsets that systematically remove sequences with training-set overlap for each competitor. As expected, the performance of CoPRA and DeePNAP declines as this overlap is reduced, whereas the performance of ZeroFold remains stable. These results demonstrate that ZeroFold achieves cutting-edge performance in protein-RNA binding affinity prediction, and that this advantage becomes more pronounced under fairer evaluation conditions.

## Results

**ZeroFold training**

To train and evaluate ZeroFold, we assembled PRADB, a curated dataset of 2,621 unique protein-RNA pairs with experimentally measured binding affinities (pKD) drawn from four complementary sources: ProNAB[29], BioLiP2[30], the UTexas Aptamer Database[31], and PDBbind+[32] (**Figure 2A**). ProNAB constitutes the largest single contribution, reflecting its broad coverage of RNA-binding protein families, while the remaining three databases provide complementary coverage of structurally characterised complexes and aptamer interactions. The 2,621 unique pairs derive from 512 distinct proteins and 1,411 distinct RNA sequences, with protein sequence lengths spanning a wide range that encompasses both compact RNA-binding domains and larger multidomain assemblies (**Figure 2C**), and RNA sequence lengths similarly covering a broad spectrum from short aptamers to longer structured RNAs (**Figure 2B**). The distribution of experimentally measured affinities ($pK_D$) is approximately unimodal and centred around 7.1, with a standard deviation of 1.35, reflecting a dataset that is neither dominated by very low-affinity nor very high-affinity binders and spans a range well suited to training a regression model (**Figure 2D**).

**ZeroFold performance**

PRADB was split into training, validation, and test sets in an approximate 8:1:1 ratio. To ensure fair evaluation and reduce potential data leakage, these splits were constructed such that no protein sequences shared greater than 40% sequence identity with at least 80% coverage across different sets. The same criterion was applied to RNA sequences.

The final model was trained for 100 epochs, with each epoch comprising 900 samples from the training set weighted by cluster representation (see Methods). The model performed strongly on the test set, achieving a mean absolute error (MAE) of 1.14, a root mean squared error (RMSE) of 1.47, a Pearson correlation coefficient (PCC) of 0.63, and a Spearman correlation coefficient (SCC) of 0.65.

Noise and variability in the experimental affinity measurements impose an effective upper bound on achievable correlation coefficients of approximately 0.6-0.7 (**Figure 2E**). This limitation reflects the quality and availability of current experimental data rather than a fundamental constraint of the modelling approach. As a result, models reporting substantially higher correlations may reflect overfitting or data leakage from the training set to the test set. The performance of ZeroFold therefore approaches the practical upper limit imposed by the quality and consistency of the available experimental data, suggesting that further improvements are likely to require larger, higher-quality, and more representative binding affinity datasets.

In addition to predictive performance, ZeroFold offers a substantial advantage in computational efficiency over structure-based affinity prediction pipelines. Both ZeroFold and structure-based approaches require a Boltz-2 trunk forward pass; however, structure-based pipelines additionally incur the cost of structure decoding prior to affinity estimation, whereas ZeroFold passes pre-structural embeddings directly to its affinity prediction head, bypassing this step (**Figure 3**). This reduction in computational overhead enables higher-throughput evaluation of protein-RNA interactions, making ZeroFold particularly well suited to applications such as virtual screening and proteome-wide affinity profiling, where the ability to evaluate large numbers of candidate sequences is critical.

**Comparison with CoPRA**

A comparative analysis against existing state-of-the-art predictors was conducted. Most current methods for protein-RNA affinity prediction are structure-based, meaning that predictions are made using experimentally determined 3D structures in which interaction information is already implicitly encoded. This provides a substantial advantage compared with approaches that rely solely on sequence information, particularly because such methods are typically evaluated on complexes with resolved structures rather than on sequence pairs lacking structural data.

The leading structure-based protein-RNA affinity predictor is CoPRA, whose reported performance substantially exceeds that of other competing methods in the field[17]. To enable a more comparable evaluation on protein-RNA pairs without experimentally resolved structures, predicted complex structures were first generated using Boltz-2 and subsequently used as input for CoPRA.

CoPRA reports a PCC of 0.58 and a SCC of 0.59, based on five-fold cross-validation of its PRA310 dataset. However, the sequence similarity threshold used to separate clusters in this benchmark was 70% on the protein sequence, which is substantially more permissive than the 40% identity threshold on both the RNA and the protein applied when constructing the ZeroFold test set. This more relaxed criterion increases the likelihood that closely related sequences appear across training and evaluation folds and, when combined with the use of experimentally resolved structural data, may contribute to inflated reported performance metrics.

To enable a fair comparison between ZeroFold and CoPRA, performance was evaluated on five progressively stricter subsets of the ZeroFold test set (**Table 1**). In the first setting (A), evaluation was performed on the full ZeroFold test set, regardless of whether any sequences overlapped with the CoPRA PRA310 dataset. In the second setting (B), any protein sequences present in both the ZeroFold test set and PRA310 were removed. In the third setting (C), protein sequences with greater than 70% sequence identity to sequences in PRA310, the similarity threshold used in the CoPRA benchmark, were excluded. In the fourth setting (D), a stricter 40% protein sequence identity threshold was applied. Finally, in the most stringent setting, both protein and RNA sequences were required to have less than 40% identity to any sequence in the training data, mirroring the criteria used to construct the ZeroFold test set and therefore providing the most comparable evaluation to the reported ZeroFold performance. This resulted in too few datapoints to make a meaningful comparison, so it is not reported in **Table 1**.

As expected, the performance of CoPRA decreased, with the PCC falling from 0.50 in A to 0.22 in D, as the evaluation criteria became more stringent, reflecting the increasing difficulty of predicting affinities without closely related sequences or experimentally resolved structures. In contrast, the performance of ZeroFold remained largely consistent across these subsets, PCC around 0.6. Notably, ZeroFold outperformed CoPRA even in the least restrictive settings, where identical or highly similar protein sequences were present in PRA310, but not in PRADB, and CoPRA therefore had the greatest potential advantage. These results demonstrate that ZeroFold achieves stronger predictive performance despite operating without experimentally determined structural inputs and under more stringent evaluation conditions.

**Comparison with DeePNAP**

A more direct comparison for ZeroFold is with other sequence-based affinity prediction methods. The current cutting-edge model of this type is DeePNAP[20], which was trained on the ProNAB dataset, a dataset that also forms part of PRADB. DeePNAP reports a correlation coefficient (R) of 0.92 for $K_D$ prediction. However, the CoPRA study reports substantially lower correlation coefficients (PCC and SCC) of 0.35 and 0.35, respectively, when evaluating DeePNAP on the PRA201 dataset. This reduction in performance likely reflects differences in dataset construction: DeePNAP does not apply a sequence similarity threshold when generating its test set, increasing the likelihood that closely related sequences appear in both training and evaluation data and thereby inflating the reported performance.

To enable a fairer comparison with ZeroFold, the same four test set subsets (A-D) used in the CoPRA analysis were constructed again, but this time using ProNAB as the reference training dataset for sequence similarity filtering. As the filtering criteria became progressively more stringent, removing identical sequences and then increasingly similar protein and RNA sequences, the performance of DeePNAP declined slightly, reflecting the increasing difficulty of predicting affinities for sequences that are less similar to those observed during training. However, this decrease was far less pronounced than the performance drop observed for CoPRA under the stricter evaluation settings. DeePNAP was trained on a considerably larger dataset, which likely contributes to its stronger generalisation across these subsets and further highlights the importance of larger training datasets for improving model performance. The performance of ZeroFold also decreased on the more stringent subsets, suggesting that these later subsets are intrinsically more challenging and that the difficulty cannot be explained solely by sequence distance from the training set. Nevertheless, ZeroFold consistently outperformed DeePNAP across all evaluated metrics.

To further characterise model performance across the affinity range, the test set was divided into low-, medium-, and high-affinity bands, defined by thresholds at the mean test-set affinity ± one standard deviation. Correlation metrics were then computed separately within each group (**Table 2**). The correlation calculated across the full test set was higher than that observed within any individual tercile. This indicates that much of the global correlation arises from between-group variance across the affinity spectrum, whereas analysing terciles in isolation removes this variance and reveals the weaker within-group ranking signal. These results suggest that the model has learned robust coarse-grained discrimination: it reliably distinguishes weak binders from strong binders. However, within-tercile correlation decreases toward the extremes of the affinity range, particularly among high-affinity interactions. In practical terms, this performance profile is well-suited to virtual screening applications, where the primary objective is to identify promising candidate binders from a large pool of sequences with unknown affinity. By contrast, the model is likely to be less reliable for fine-grained lead optimisation tasks, with an SCC of only 0.28 in the stronger binders, where subtle differences in affinity between already potent binders must be resolved. It is also important to note that ZeroFold has a higher SCC than both CoPRA and DeePNAP across all bands.

**Discussion**

We have reported how ZeroFold achieves state-of-the-art performance in protein-RNA binding affinity prediction, with accuracy approaching the practical upper bound imposed by experimental noise and inter-assay variability[29-32]. The proximity of model performance to the level of inter-measurement variability suggests that current data quality may constitute a major limiting factor, and that further gains may require larger and more consistently measured training datasets[29-32], and that further gains will likely require larger and more consistently measured training data rather than architectural improvements alone.

A central finding is that the advantage of ZeroFold over both structure-based and sequence-based baselines widens as evaluation conditions become fairer. Against CoPRA, its lead grows as test subsets are progressively filtered to remove sequences with training-set overlap, and this holds even though CoPRA has access to experimentally resolved structures[17]. Against DeePNAP, ZeroFold outperforms consistently across all filtered ProNAB subsets despite operating on a smaller training set[20]. Together, these comparisons indicate that the gains of ZeroFold reflect genuine generalisation rather than favourable dataset composition.

These results support the view that pre-structural embeddings are a particularly informative representation for modelling interactions involving conformationally heterogeneous partners[26,27]. By extracting intermediate representations from the Boltz-2 trunk before the structure decoding step, ZeroFold retains ensemble-level information that is discarded when a dynamic system is collapsed into a single predicted conformation[26,28]. This is especially consequential for protein-RNA systems, where RNA flexibility, context dependence, and binding-induced rearrangements limit the utility of static structural models[14]. Crucially, the approach requires no resolved structure at inference time, making it directly applicable to the large majority of protein-RNA pairs for which structural data do not exist.

Several limitations remain. The available quantitative protein-RNA affinity data are modest in scale relative to those underpinning advances in other areas of machine learning for biology, and the integration of measurements from different laboratories and experimental modalities may obscure finer-grained biophysical trends. Prediction accuracy is also likely to vary across protein classes, RNA types, and affinity regimes, and future work should characterise these differences systematically. It will additionally be important to test whether the pre-structural embedding strategy generalises to related tasks such as mutation-effect prediction, binding-site identification, and the modelling of multicomponent ribonucleoprotein assemblies.

In summary, ZeroFold demonstrates that pre-structural embeddings offer a general strategy for learning interaction energetics in systems where conformational flexibility is central to function. As the performance is approaching the ceiling set by current data quality, progress beyond this point will particularly depend on the generation of larger, more reproducible affinity datasets.

**Methods**

ZeroFold comprises four main components: the RNA encoder, the protein encoder, the transformer layers and the affinity prediction head.

**RNA/Protein input representations**. ZeroFold employs Boltz-2[28] to extract representations for both protein and nucleic acid chains. These representations were extracted from the final layer of the Boltz-2 trunk (recycling_steps 3), just before the structure module decoding. The protein and nucleic acid representations each consist of two components. Firstly, we extracted the single sequence representation $\{s_i^{msa}\}$, where $s_i^{msa} \in R^{c_{msa}}, c_{msa} = 384$, and $i \in \{1 \ldots N_{res}\}$, which is the per-residue (or per-nucleotide) embedding from the final trunk layer. Secondly, we extracted the pair representations, $\{z_{ij}^{pair}\}$, where $z_{ij}^{pair} \in R^{c_{pair}}, c_{pair} = 128$ and $i, j \in \{1 \ldots N_{res}\}$, which are the pairwise embeddings encoding residue–residue (or nucleotide–nucleotide) spatial and evolutionary relationships.

**RNA-protein encoder**. The protein encoder maps Boltz-2 trunk representations into refined single and pair feature streams. The single representation $s_i^{msa}$ and pair representation $z_{ij}^{pair}$ are each passed through two successive transition layers with residual connections and dropout, preserving the input dimensionality throughout. The RNA encoder follows an identical structure, with the addition of a learned nucleic acid type embedding (RNA vs DNA, 32-dimensional) concatenated to the single representation before the first transition layer (which, due to the resulting dimension change from 384 to 416, is applied without a residual connection), allowing the encoder to condition on nucleic acid type.

**Cross-modal attention module and affinity head**. The cross-modal attention module integrates the encoded protein and RNA representations to model interactions across the protein-RNA interface. The resulting joint representation is then passed to an affinity prediction head to produce a scalar $pK_D$ estimate.

**Datasets**
*ProNAB*. The primary dataset used to train ZeroFold was ProNAB, a manually curated database containing over 20,000 experimentally measured binding affinities for protein-DNA (14,606) and protein-RNA (5,323) complexes[29]. Among the protein-RNA entries, 5,173 contained valid $K_D$ measurements, which were converted to $pK_D$ values for training. After further filtering to retain only valid sequences, specifically complexes containing unmodified nucleotide bases and natural amino acids, the dataset comprised 3,588 protein-RNA pairs with valid affinity values. Many protein-RNA sequence pairs appeared multiple times in the dataset with substantially different reported affinity values (**Figure 2E**). These discrepancies likely arise from differences in experimental conditions, including temperature, pH, and assay methodology, which are extensively annotated within the ProNAB database. However, stratifying the data by these experimental variables did not substantially reduce the variability in reported $pK_D$ values, suggesting that much of the observed variation reflects measurement noise and experimental uncertainty rather than systematic differences in the annotated conditions. Because it was not possible to define a consistent set of standard experimental conditions, and because each measurement still represents a valid experimental observation of the same interaction under different contexts, all entries were retained to preserve the available experimental evidence and allow the model to learn from the full range of reported binding affinities.

**BioLiP2**. The next largest source of affinity data was BioLiP2[30]. This database contains biologically relevant protein-ligand interactions derived from the Protein Data Bank (PDB), annotated with binding affinity measurements collected from external databases and manual literature curation. BioLiP2 contains 425 protein-RNA complexes with reported binding affinity data.

***UTexas aptamer database***. The aptamer database curated by the University of Texas compiles binding affinity measurements reported in the literature between 1990 and 2022[31]. Within this resource, 258 protein-RNA complexes were associated with labelled binding affinity values.

***PDBbind+***. The final dataset used was PDBbind+ (v2020), which was considered the most reliable source of affinity data due to its high level of curation and the availability of experimentally determined 3D structural information alongside binding affinity annotations[32]. This dataset contains 239 protein-RNA complexes with reported binding affinities.

***PRADB***. The union of the four previously described databases was taken to produce a redundant set of 4,510 protein-RNA pairs. For pairs associated with multiple affinity measurements, the value reported in PDBbind+ was preferentially selected due to its higher level of curation; if no PDBbind+ measurement was available, the median affinity across the remaining sources was used. After resolving these redundancies, the final Protein-RNA Affinity DataBase (PRADB) comprised 2,621 unique protein-RNA sequence pairs. The distributions of protein sequence lengths, RNA sequence lengths, and affinity values are shown in **Figure 2B-D**.

**Clustering.** To prevent data leakage across train, validation, and test splits, all protein and nucleic acid sequences were clustered independently using MMseqs2 easy-cluster[33], with a 40% sequence identity threshold and 80% coverage requirement. This yielded 347 protein clusters and 1,151 nucleic acid clusters, with each data point assigned both a protein cluster label and a nucleic acid cluster label. Split assignment was then performed at the cluster level, where no cluster representative could appear in more than one split, ensuring that no sequence in the evaluation sets shares more than 40% identity with any training sequence. Following an initial split assignment, the splits were audited using bidirectional MMseqs2 easy-search to detect leakage cases that greedy clustering can miss: two sequences above the identity threshold can occupy different clusters if each is more similar to a different representative. This search revealed 23 entries in the validation and test sets with sequence identities exceeding 40% to training counterparts; these were removed, yielding final split sizes of 2,104 training, 210 validation, and 299 test data points.

**Sample weighting**. To counteract the over-representation of highly sampled sequence families in the training set, we follow a similar protocol to AlphaFold-Multimer, where each data point was assigned a sample weight inversely proportional to the size of its protein and nucleic acid clusters:

$$w_{ij} = \frac{1}{|PC_i|} \cdot \frac{1}{|RC_j|}$$

where $PC_i$ is the protein cluster $i$ and $PC_j$ is the RNA cluster $j$. Weights were normalised so that they sum to unity across the training set. This scheme down-weights complexes from densely sampled families (such as ribosomal proteins paired with rRNA) and up-weights unique or rare sequence pairs, encouraging the model to learn from the full diversity of the dataset rather than fitting the dominant clusters.

**Funding.** This work was supported by AstraZeneca through a PhD studentship. Part of this work was performed using resources provided by the Cambridge Service for Data Driven Discovery (CSD3) operated by the University of Cambridge Research Computing Service (www.csd3.cam.ac.uk).

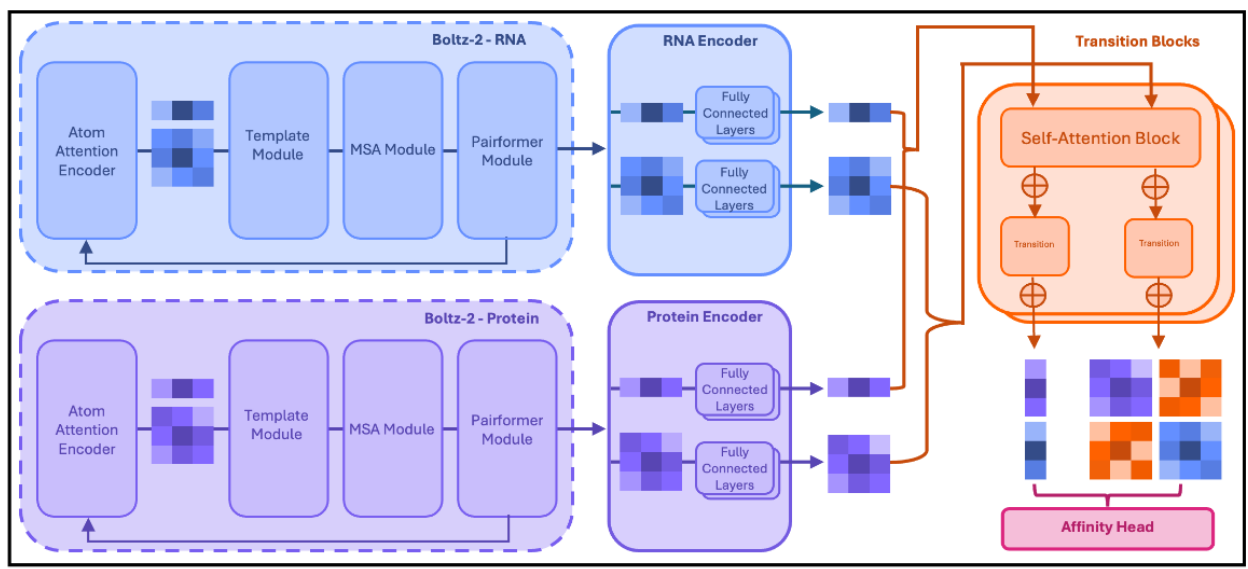

**Figure 1. Architecture of ZeroFold.** ZeroFold predicts protein-RNA binding affinities from sequences by operating on pre-structural embeddings extracted from the trunk of Boltz-2 prior to its structure decoding step. These embeddings encode rich conformational ensemble information for both the protein and RNA chains, and they are used to construct a heterogeneous graph in which protein residues and RNA nucleotides form nodes and pairwise relationships form edges. Separate encoding blocks refine the single and pair representations for each chain before a cross-modal attention module integrates information across the protein-RNA interface. The resulting joint representation is passed to an affinity prediction head to produce a scalar binding affinity estimate. Components enclosed in dotted-line boxes have fixed parameters and are not updated during training; those enclosed in solid-line boxes are trainable. The Boltz-2 trunk is fixed throughout, with ZeroFold trained on the downstream encoding, attention, and prediction components.

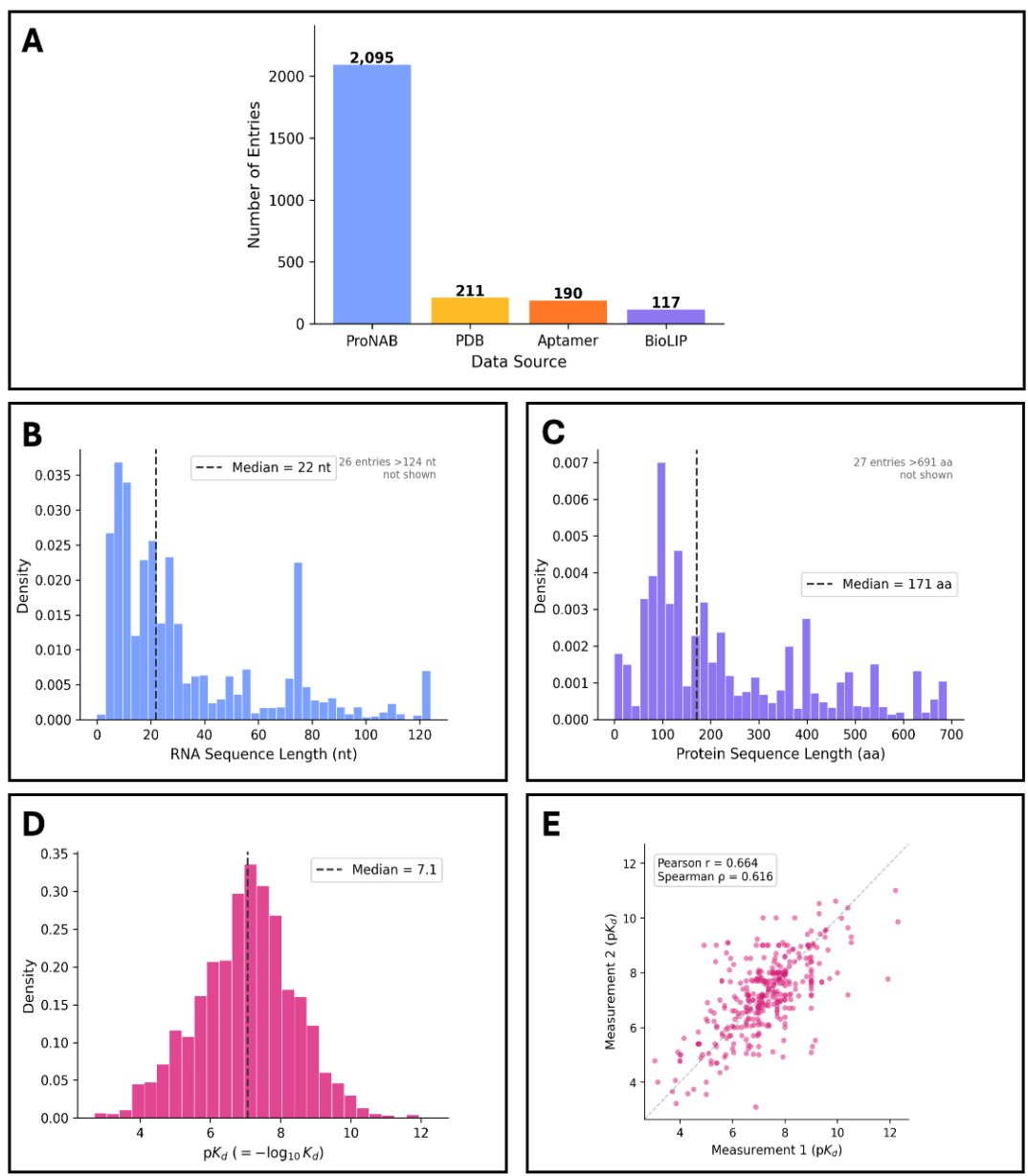

**Figure 2. Composition and characteristics of PRADB. (A)** Breakdown of the 2,621 unique protein-RNA pairs in PRADB by source database, illustrating the dominance of ProNAB and the complementary coverage provided by BioLiP2, the UTexas Aptamer Database, and PDBbind+. **(B)** Distribution of RNA sequence lengths across 1,411 unique RNA sequences, reflecting the broad range of RNA classes represented in the dataset. **(C)** Distribution of protein sequence lengths across 512 unique proteins, highlighting the diversity of RNA-binding protein families included. **(D)** Distribution of experimentally measured binding affinities ($pK_D$) across all 2,621 protein-RNA pairs, showing the range and central tendency of affinity values used for model training and evaluation. **(E)** Agreement between independent replicate affinity measurements for the 353 protein-RNA pairs with two or more experimental values, illustrating the degree of experimental noise that sets a practical upper bound on achievable model performance.

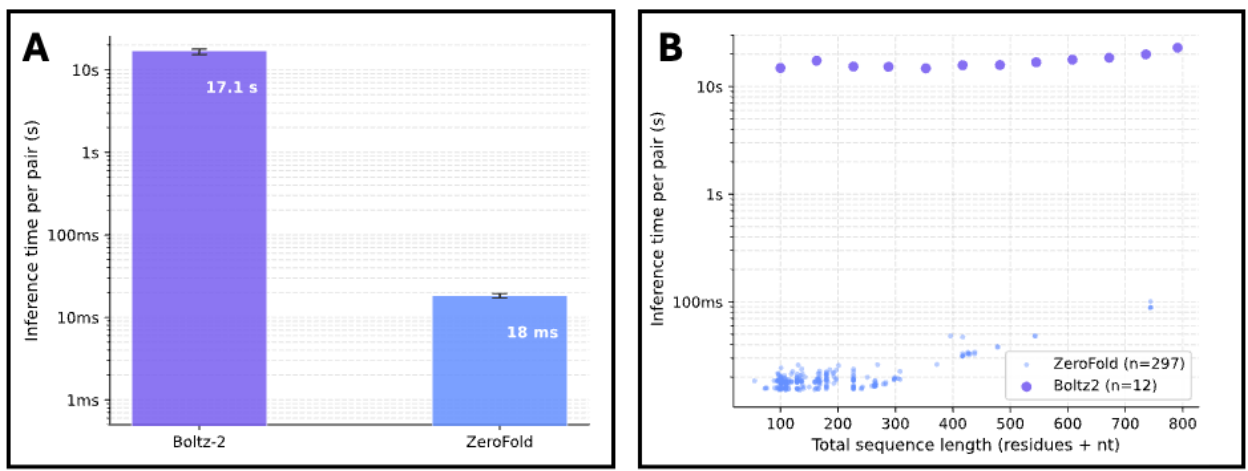

**Figure 3. Computational efficiency of ZeroFold compared with Boltz-2.** Inference time per protein-RNA pair for ZeroFold and a structure-based pipeline in which Boltz-2 is used to generate predicted complex structures, which are then used as input for affinity estimation. Both pipelines require a Boltz-2 trunk forward pass; the structure-based pipeline additionally performs structure decoding prior to affinity estimation, whereas ZeroFold passes pre-structural embeddings directly to its affinity prediction head, bypassing this step. Bars represent mean inference time per sample; error bars indicate standard deviation across independent runs. The reduction in runtime achieved by ZeroFold reflects the elimination of the structure decoding step, enabling higher-throughput evaluation of protein-RNA pairs compared with structure-based approaches.

|  | vs CoPRA (PRA310) | | | | vs DeePNAP (ProNAB) | | | |
|  | Pcc | | Scc | | Pcc | | Scc | |
| Subset | n | CoPRA | ZeroFold | CoPRA | ZeroFold | n | DeePNAP | ZeroFold | DeePNAP | ZeroFold |
| --- | --- | --- | --- | --- | --- | --- | --- | --- | --- | --- |
| A) All | 297 | 0.50 | **0.63** | 0.54 | **0.65** | 297 | 0.46 | **0.63** | 0.35 | **0.65** |
| B) Protein not in reference | 280 | 0.47 | **0.67** | 0.52 | **0.68** | 57 | 0.48 | **0.54** | 0.42 | **0.54** |
| C) Max protein identity < 70% | 162 | 0.46 | **0.71** | 0.55 | **0.69** | 32 | 0.44 | **0.61** | 0.45 | **0.56** |
| D) Max protein identity < 40% | 97 | 0.22 | **0.54** | 0.23 | **0.52** | 29 | 0.41 | **0.53** | 0.34 | **0.46** |

**Table 1. ZeroFold performance vs CoPRA and DeePNAP. (A-D)** Comparison of ZeroFold performance on the full test set (A) and progressively stricter subsets (B-D) against CoPRA and DeePNAP. These subsets impose increasingly stringent constraints on sequence overlap between the subsets and the training datasets of CoPRA (PRA310) and DeePNAP (ProNAB).

|  |  | Spearman Correlation Coefficient | | |
| Affinity Band | n | CoPRA | DeePNAP | ZeroFold |
| --- | --- | --- | --- | --- |
| Low (true < 5.75) | 48 | 0.09 | 0.14 | 0.10 |
| Medium (5.75 − 8.45) | 198 | 0.38 | 0.02 | **0.50** |
| High (true ≥ 8.45) | 51 | 0.27 | 0.20 | **0.28** |

**Table 2. Performance of models on stratified test set.** Correlation of models on the test set stratified by affinity band. Band boundaries are defined as the test-set mean ±1 standard deviation (mean = 7.10, SD = 1.35; low < 5.75, medium 5.75-8.45, high ≥ 8.45).